\documentstyle{elsart}

\begin{document}
\begin{frontmatter}
\title{Coexistence of Coherence and Incoherence in Nonlocally Coupled
Phase Oscillators:\\
A Soluble Case}
\author[Kyoto]{Yoshiki Kuramoto}
\author[US]{Dorjsuren Battogtokh}
\address[Kyoto]{Department of Physics, Graduate School of Sciences, 
Kyoto University,Kyoto 606-8502, Japan}
\address[US]{Department of Physics and Astronomy, Louisiana State University,
Baton Rouge, LA 70803-4001, USA}
\begin{abstract}
The phase oscillator model with global coupling is extended
to the case of finite-range nonlocal coupling. Under suitable conditions, 
peculiar patterns emerge in which a quasi-continuous array of identical
oscillators separates sharply into two domains,
one composed of mutually 
synchronized oscillators with unique frequency 
and the other composed of desynchronized
oscillators with distributed frequencies. We apply a theory similar to the one
which successfully explained the onset of collective synchronization
in globally coupled phase oscillators with frequency distribution. A 
space-dependent order parameter 
is thus introduced, and an exact functional self-consistency equation is 
derived
for this quantity. Its numerical solution is confirmed to reproduce the 
simulation results accurately.
\end{abstract}
\end{frontmatter}

\section{Introduction}
Large populations and continuous fields of coupled oscillators form a
representative class of synergetic systems met in a wide range of 
scientific disciplines 
from physics, chemistry, engineering, biology to 
brain science \cite{ref1,ref2,ref3,ref4,ref5}.
Collective dynamics of coupled oscillators depends crucially on the range of 
their mutual coupling. It has recently been realized that when the coupling is 
nonlocal, the patterns which emerge could be drastically different
from those which we expect for oscillators with local or global coupling
\cite{ref6,ref7,ref8}. 
The implication of this fact is relevant even to what
we conventionally call locally coupled systems, 
typically reaction-diffusion systems. This is because it may happen that 
nonlocality can arise effectively as a result of elimination of some variables, e.g., rapidly diffusing components in the case of reaction-diffusion dynamics. 
Among the variety of patterns which are characteristic to nonlocally coupled 
oscillators, we will focus our attention below on
a particular class of patterns in which the whole medium is separated
into two 
domains of qualitatively different dynamics. Specifically, the oscillators are 
mutually synchronized 
in one domain while they are completely desynchronized.
in the other domain. A preliminary work on such dynamics
was reported recently\cite{ref8}. We 
will present below a more thorough investigation of this problem.

The collective dynamics of our concern is similar to the collective 
synchronization in globally coupled oscillators with distributed natural 
frequencies \cite{ref4,ref9,ref10} 
where the whole population splits into two subpopulations 
each composed of synchronized and desynchronized oscillators. There
the systems is stationary in a statistical sense within a constant drift of
the collective phase corresponding to the oscillation of the 
population as a whole.
Similarly to the case of global coupling,
the individual oscillators can be regarded as being
controlled by a common mean field, although the latter is
now space dependent. 
The oscillators under consideration are 
identical in nature, still
we may develop an exact mean-field theory similar
to the one developed for globally coupled oscillators.
Commonly to these theories,
a suitably defined order parameter representing the collective state
is determined from a condition
for self-consistency
which must exist between the collective dynamics 
and the dynamics of the individual oscillators. 

In section 2, some results of our numerical simulation will be presented.
The simulation is carried out first on a nonlocal version of the complex 
Ginzburg-Landau (CGL) equation in one space dimension with periodic boundary
conditions, and then on another equation obtained through the phase reduction
of the nonlocal CGL.
In section 3, our theory will be developed and a functional self-consistency
equation will be derived for the order parameter. Then the numerical solution
of the self-consistency equation will be compared with the simulation results
presented in Section 2. 
Concluding remarks will be given in the final section.
\section{Coexistence of coherence and incoherence} 
As a simplest model for densely and uniformly distributed oscillators with 
nonlocal coupling, let us consider the following equation for a complex
amplitude $A$ which we call
nonlocally coupled complex Ginzburg-Landau equation\cite{ref5,ref6}. 
In one spatial
dimension our model is expressed as
\begin{equation}
\frac{\partial}{\partial t}A(x,t)=(1+i\omega_0)A-(1+ib)|A|^{2}A
+K(1+ia)\Bigl( Z(x,t)-A(x,t) \Bigr).
\end{equation}
The quantity $Z(x,t)$, which we call the mean field, represents the effects of
the nonlocal coupling and is given by
\begin{equation}
Z(x,t)=\int G(x-x')A(x',t)dx'.
\end{equation}
The coupling function $G$ changes with the distance as 
\begin{equation}
G(y)=\frac{\kappa}{2}\exp(-\kappa |y|)
\end{equation}
and is normalized. The exponential form of $G$ is in fact a natural consequence
of the reductive derivation of Eq.\ (1) from a certain class of 
reaction-diffusion systems when the latter involve an inactive diffusive 
component to be eliminated 
adiabatically\cite{ref6,ref7}.
Note that the last term in Eq.\ (1) representing the coupling is 
so arranged
that it may vanish when $A$
is uniform in space. If $A$ is nonuniform but sufficiently long-waved, this 
term can be approximated with a diffusion term, and then 
Eq.\ (1) is reduced to the ordinary CGL. 

Equation (1) is numerically simulated on a finite interval $[0,1]$ 
with periodic boundary
conditions. In a suitable range of parameters, in which the uniform oscillation
is linearly stable, the system can develop peculiar
patterns when the initial perturbation is finite. Figure 1
shows a typical example of such phase patterns. 
The initial condition is such that
the modulus of $A$ equals 1 everywhere and its phase changing randomly
still its envelope being nearly symmetric about the midpoint $x=1/2$.
Since the coupling constant $K$ is assumed to be
relatively small, the local oscillations attain almost full amplitude, 
so that we will mainly be interested in the phase pattern.
The pattern consists of 
two distinct
domains. In one domain which appears near the boundaries, the pattern
is spatially continuous and smooth, whereas in the central domain 
spatial
continuity seems completely lost.
The pattern as a whole looks steadily advancing upward, implying that
the entire population is undergoing regular collective 
oscillation with a definite frequency $\Omega$.

To investigate the nature and origin of such patterns in further detail, 
Eq.\ (1) is reduced to a phase equation which would be much easier 
to analyze. 
The phase reduction is valid when the coupling strength $K$ is small,
which is actually the case, and the reduced equation takes the form
\begin{equation}
\frac{\partial}{\partial t}\phi(x,t)=\omega-\int G(x-x')\sin \Bigl(
\phi(x,t)-\phi(x',t)+\alpha \Bigr)dx',
\end{equation}   
where the time scale has been so changed as to normalize the coupling strength,
$\omega$ is the natural frequency rescaled accordingly,
and the phase constant $\alpha$ in the phase-coupling function is related to
the original parameters through
\begin{equation}
\tan\alpha=\frac{b-a}{1+ab},\quad \alpha(b-a)>0.
\end{equation}
Under the parameter condition corresponding to the previous
numerical simulation of Eq.\ (1), a phase pattern similar to FIG.\ 1 
was obtained, which is displayed
in FIG.\ 2. 

We now focus on the phase patterns obtained for our phase equation.
Let us introduce relative phase $\psi$ by
\begin{equation}
\phi=\Omega t+\psi
\end{equation}
which describes the dynamics of the phase deviation from the reference motion
with some drift velocity $\Omega$ whose value is still open. 
We rewrite Eq.\ (4) using $\psi$ as
\begin{equation}
\frac{\partial}{\partial t}\psi(x,t)=\omega-\Omega-\int
dx' G(x-x')\sin\Bigl( \psi(x,t)-\psi(x',t)+\alpha \Bigr).
\end{equation}
As a generalization of the theory of synchronization transition of globally
coupled phase oscillators with frequency distribution, we
introduce a complex order parameter with modulus $R$
and phase $\Theta$ through
\begin{equation}
\int dx' G(x-x')\exp[i\psi(x',t)]=R(x,t)\exp[i\Theta(x,t)].
\end{equation}
Unlike the order parameter defined for globally coupled oscillators,
the above quantity is space-dependent. Still a physical picture is
valid such that
we are practically working with an assembly of {\em independent} oscillators 
under the 
control of a common forcing field represented by $R$ and $\Theta$.
This is confirmed by rewriting Eq.\ (7)
in terms of the order parameter into a forced one-oscillator equation as
\begin{equation}
\frac{\partial}{\partial t}\psi(x,t)=\omega-\Omega-R(x,t)\sin
\Bigl( \psi(x,t)+\alpha+\Theta(x,t) \Bigr).
\end{equation}
The spatial profiles of $R(x)$ and $\Theta(x)$ after a long-time average 
are displayed in FIG.\ 3a and 3b, respectively.
We see that the forcing mean-field amplitude is stronger near the 
boundaries and weaker near the
center of the system. 
If the phase distribution such
as shown in FIG.\ 2 is statistically stationary within a constant overall 
drift with
velocity $\Omega$, then $R$ and $\Theta$ are time-independent.
Their approximate constancy 
is confirmed by the fact that the fluctuation of the time sequence of $R(x,t)$
observed at the midpoint of the system $x=1/2$ gives a standard
deviation as small as 0.008 which comes possibly from a finite-size effects.
Similar property can also be confirmed for $\Theta(x,t)$.

Figure 3c shows a
distribution of the actual frequencies $\bar{\omega}(x)$ 
of the individual oscillators.
This was also obtained through a long-time average. In the coherent domain,
the oscillation frequencies have an identical value $\Omega$,
while in the incoherent domain
they are distributed but give a well-defined continuous curve.

The dynamics which is going on
is now clear. The system is divided into two subgroups of 
oscillators. In the first group, 
the forcing amplitude is large enough for the oscillators to be entrained, so 
that they
oscillate with an identical frequency $\Omega$. In the second
group, in contrast, the forcing amplitude is too weak for entrainment, so that
the frequencies of the individual oscillators differ from $\Omega$. 
Since the actual frequencies of these desynchronized oscillators
 should depend on the local amplitude of the forcing,
the nonuniform but smooth pattern of $R(x)$ as shown in FIG.\ 3a implies a
smooth distribution of the actual frequencies, and this is consistent with
FIG.\ 3c showing $\bar{\omega}(x)$.
Our next problem is how to predict stationary patterns of $R(x)$ and 
$\Theta(x)$
theoretically, and this will be the subject of the next section.
\section{Theory}
The basic equations to work with below are Eqs.\ (8) and (9).
Assuming $R$ and $\Theta$ to be time-independent, we first try to find 
solutions
of Eq.\ (9) for the phases $\psi(x,t)$ each of which should be a function of 
$R(x)$ and $\Theta(x)$. These solutions are then substituted into Eq.\ (8),
leading to an equation for determining $R$ and $\Theta$ as a {\em functional}
of $R$ and $\Theta$ in a self-consistent manner. Since the collective 
frequency $\Omega$ is still to be determined, we will be working with 
a nonlinear eigenvalue problem.

Here a question may arise as to the seeming contradiction between the 
assumed time-independence of $R$ and $\Theta$ and the apparent time-dependence
of the solution of Eq.\ (9) which is true if $R(x)$ is 
below a certain
critical value.
The same problem was encountered in the problem of synchronization transition
in globally coupled phase oscillators with frequency distribution.
The answer to this question is that the factor
$\exp[i\psi(x',t)]$ appearing in the integral of Eq.\ (8) should be
replaced with is statistical
average $\langle \exp[i\psi(x',t)] \rangle$. Such an average can be
 calculated using a suitable
invariant measure which is easy to find. 

The successive steps to be taken for reaching the final solution 
are the following. 
Note first that Eq.\ (9) admits two types of
solutions depending on the location of the oscillators of concern.
 They are constant solutions and 
drifting solutions.
If $|R/(\omega-\Omega)|\le 1$, we have a time-independent solution
given by
\begin{equation}
\theta_{0}(R)=\mbox{Sin}^{-1}\Biggl( \frac{\omega-\Omega}{R(x)}\Biggr)-\alpha.
\end{equation}
If $|R/(\omega-\Omega)|> 1$, in contrast, we have a drifting solution.
The actual frequency $\bar{\omega}$ of the corresponding desynchronized
oscillator 
as a function or $R$ is given by
\begin{equation}
\bar{\omega}=\Omega+(\omega-\Omega)
\sqrt{1-\Biggl(\frac{R}{\omega-\Omega}\Biggr)^{2}},
\end{equation}
which equals the collective frequency $\Omega$ plus the mean drift
velocity of $\psi$.
Instead of using such drifting solutions in Eq.\ (9), 
we use an invariant measure,
i.e., the probability density $p(\theta,R)$ such that $\psi$ takes on value 
$\theta$ for given $R$,
which is inversely proportional to the drift velocity of $\psi$ at
$\psi=\theta$. Thus,
\begin{equation}
p(\theta,R)=C(\omega-\Omega-R\sin\theta)^{-1},
\end{equation} 
where C is a normalization constant given by $C=(2\pi)^{-1}
\sqrt{(\omega-\Omega)^{2}-R^{2}}$. As noted earlier, this invariant measure
is used for replacing the exponential factor
appearing on the left-hand side of Eq.\ (8) with its statistical average.
The physical idea underlying this replacement is that an equilibrium
statistical ensemble may be applicable to each small local subsystem
because infinitely many oscillators which are independently oscillating
are already contained in such a subsystem.

The spatial profile of $R(x)$ in FIG.\ 3 implies that
there are spatial points $x=\frac{1}{2}\pm x_{c}$ at which $R$ takes a
critical value separating the two types of solutions described above.
Namely, in the inner domain ($|x'-\frac{1}{2}|\ge x_{c}$) the
solutions are drifting, while in the outer domain
 ($|x'-\frac{1}{2}|\ge x_{c}$) they are constant in time.
Thus Eq.\ (8) becomes
\begin{equation}
\int_{0}^{1}dx'G(|x-x'|)\exp\Bigl[i\Bigl(\Theta(x')-\alpha\Bigr)\Bigr]
h\Bigl(R(x')\Bigr)=R(x)\exp[i\Theta(x)],
\end{equation}
where 
\begin{equation}
h(R)=\left\{\begin{array}{ll}
\exp[i\theta_{0}(R)] &  \quad (|x'-\frac{1}{2}|\ge x_{c})\\
\int_{0}^{2\pi}\exp(i\theta)p(\theta,R)d\theta & \quad (|x'-\frac{1}{2}|< x_{c})
\end{array}\right.
\end{equation}
The above expression for $h(R)$ in each domain can further be made explicit as 
\begin{eqnarray}
\exp[i\theta_{0}(R)]&=&\sqrt{1-\Bigl( \frac{\omega-\Omega}{R} \Bigr)^2}
+i\frac{\omega-\Omega}{R},\\
\int_{0}^{2\pi}\exp(i\theta)p(\theta,R)\d\theta
&=&\frac{i}{R}\Bigl\{ \omega-\Omega-
\sqrt{(\omega-\Omega)^{2}-R^{2}}
\Bigr\}.
\end{eqnarray}
Equations (13)$\sim$(16) constitute a functional self-consistency condition
from which $R(x)$, $\Theta(x)$ and $\Omega$ are to be determined. By using
$R(x)$ and $\Omega$ thus obtained, the actual frequencies of the desynchronized oscillators
are determined from Eq.\ (11).

There is a unique value $\Omega$ of the collective frequency
for which our functional 
self-consistency equation admits a solution. Such solution can be obtained
numerically, and is compared with our numerical simulation. Some of the results
are given in FIG.\ 4a$\sim$4c. We find that the agreement between the 
theory and numerical experiments is almost perfect for each quantity of
$R(x)$, $\Theta(x)$, $\Omega$ and $\bar{\omega}(x)$. 
\section{concluding remarks}
The present theory makes full use of the fact that the number of oscillators
contained in the range of coupling is practically infinite. This fact allows
us to apply a mean field theory, and our problem is reduced to a 
one-oscillator problem supplemented with a self-consistency condition.
It is well known that the same idea works nicely for globally coupled 
oscillators.
The present theory suggests that, beyond a special problem discussed in the
present paper, the mean-field idea may work more generally for 
nonlocally coupled systems as far as the coupling radius
remains much longer than the minimum length
scale associated with the discreteness of the spatial distribution
of the constituent oscillators.

{\bf Acknowledgement}

The present work gives a slight extension of the my old work on
globally coupled phase oscillators which was developed in the middle of
1970's, shortly after which I spent a very fruitful year at Professor 
Hermann Haken's
Institute in Stuttgart. Therefore, it is my great pleasure to dedicate
this small work to Professor Haken. 

\newpage
\begin{center}
Figure Captions
\end{center}
\begin{itemize}
\item [FIG.1.] Instantaneous spatial distribution of the phases obtained from
Eq.\ (1) approximated with an array of 512 oscillators distributed
over an interval of length 1 with periodic boundary conditions. 
Parameter values are 
$a=-1.0$, $b=0.88$, $\kappa=4.0$.

\item [FIG.2.] Instantaneous spatial distribution of the phases obtained from
Eq.\ (4) approximated with an array of 512 oscillators distributed over
an interval of length 1 with periodic boundary conditions.
Parameter values are $\alpha=1.457$, $\kappa=4.0$.

\item [FIG.3.] Spatial profiles of the long-time average of the 
order parameter amplitude $R(x)$ (a), the order parameter phase
$\Theta(x)$ (b) and actual frequencies $\Omega(x)$ (c), each obtained
from Eq.\ (4).  
 
\item [FIG.4.] Theoretical curves
for 
$R(x)$ (solid line in (a)), $\Theta(x)$ (solid line in (b)) and $\Omega(x)$ 
(broken line in (c)) are compared with the corresponding 
quantities in FIG.\ 3 obtained numerically. 
\end{itemize}
\end{document}